\documentclass[12pt]{iopart}

\usepackage{graphicx,setspace}
\usepackage{dsfont}
\usepackage{iopams}
\usepackage{mathrsfs}
\usepackage{psfrag}
\usepackage{cases}

\begin{document}
\title{Canonical quantization of electromagnetism in spatially dispersive media}
\author{S A R Horsley and T G Philbin}
\address{Physics and Astronomy Department, University of Exeter,
Stocker Road, Exeter EX4 4QL, UK.}
\eads{\mailto{s.horsley@exeter.ac.uk}, \mailto{t.g.philbin@exeter.ac.uk} }

\begin{abstract}
We find the action that describes the electromagnetic field in a spatially dispersive, homogeneous medium. This theory is quantized and the Hamiltonian is diagonalized in terms of a continuum of normal modes.  It is found that the introduction of nonlocal response in the medium automatically regulates some previously divergent results, and we calculate a finite value for the intensity of the electromagnetic field at a fixed frequency within a homogeneous medium.  To conclude we discuss the potential importance of spatial dispersion in taming the divergences that arise in calculations of Casimir--type effects.
\end{abstract}


\section{Introduction}
\par
In macroscopic electromagnetism the dependence of the dielectric functions on frequency, and the relation of this dependence to absorption, have a central place. The circumstances where dispersion and absorption can be neglected are a small subset of cases where macroscopic electromagnetism is successfully applied. In contrast, the dependence of the dielectric functions on wavevector (spatial dispersion, or nonlocal response) is normally regarded as of rather limited relevance (see e.g.~\cite{LLcm}).  Nevertheless, there are some situations where the inclusion of spatial dispersion is required in order to obtain accurate results.  Recent work in the nano--optics of small metallic particles~\cite{raz11,wie12,pen12} has found that a nonlocal description can have a significant---and sometimes counterintuitive---effect on the localization of radiation.  For example, in the system of touching nanowires investigated in~\cite{pen12} the inclusion of spatial dispersion turns a continuous spectrum into a discrete one.  In this work we are interested in the consequences of a nonlocal description for the \emph{quantum mechanical} properties of the electromagnetic field.  We shall demonstrate that these are also nontrivial, and that a nonlocal description is sometimes necessary in order to obtain finite results.
\par
Here we aim both to set up a canonical formulation of quantum electromagnetism in spatially and temporally dispersive media (in this respect we look to extend~\cite{phi10} and add to the work of Scheel and coworkers~\cite{buh2012}, and Suttorp~\cite{suttorp2007}), and to show that such an extension to the theory can naturally cure some divergences that plague the local theory.  The simplest example of such a divergence is evident in the zero temperature electric field correlation function in a homogeneous medium, written in terms of the electromagnetic Green function, \(\boldsymbol{G}\),
\begin{equation}
	\fl\langle0|\hat{\boldsymbol{E}}(\boldsymbol{r},t)\boldsymbol{\otimes}\hat{\boldsymbol{E}}(\boldsymbol{r}^{\prime},t)|0\rangle=\int_{0}^{\infty}d\omega\boldsymbol{c}(\boldsymbol{r},\boldsymbol{r}^{\prime},\omega)=\frac{\hbar\mu_{0}}{\pi}\int_{0}^{\infty}d\omega\omega^{2}\rm{Im}\left[\boldsymbol{G}(\boldsymbol{r},\boldsymbol{r}^{\prime},\omega)\right]\label{correlation-function},
\end{equation}
It is well known that equation  (\ref{correlation-function}) diverges as \(\boldsymbol{r}\to\boldsymbol{r}^{\prime}\), even in free space due to the integration over the zero-point fields at all frequencies. However the integrand, \(\boldsymbol{c}(\boldsymbol{r},\boldsymbol{r}^{\prime},\omega)\), should certainly not diverge in this limit for it represents the intensity of the electric field at a fixed frequency, which determines single--frequency phenomena such as spontaneous emission rates. In vacuum   \(\boldsymbol{c}(\boldsymbol{r},\boldsymbol{r}^{\prime},\omega)\) is finite, but inside a dielectric it diverges. Given that the electromagnetic Green function satisfies
\[
	\boldsymbol{\nabla}\boldsymbol{\times}\boldsymbol{\nabla}\boldsymbol{\times}\boldsymbol{G}(\boldsymbol{r},\boldsymbol{r}^{\prime},\omega)-\frac{\omega^{2}\varepsilon(\omega)}{c^{2}}\boldsymbol{G}(\boldsymbol{r},\boldsymbol{r}^{\prime},\omega)=\boldsymbol{\mathds1}_{3}\delta^{(3)}(\boldsymbol{r}-\boldsymbol{r}^{\prime}) ,
\]
it has a longitudinal part
\begin{equation}
	\fl\boldsymbol{G}_{\parallel}(\boldsymbol{r},\boldsymbol{r}^{\prime},\omega)=-\frac{c^{2}}{\omega^{2}\varepsilon}\boldsymbol{\delta}_{\parallel}(\boldsymbol{r}-\boldsymbol{r}^{\prime})\to\rm{Im}[\boldsymbol{G}_{\parallel}(\boldsymbol{r},\boldsymbol{r}^{\prime},\omega)]=\frac{c^2\rm{Im}[\varepsilon]}{\omega^{2}|\varepsilon|^{2}}\boldsymbol{\delta}_{\parallel}(\boldsymbol{r}-\boldsymbol{r}^{\prime})\label{long-G}
\end{equation}
which contributes an infinite amount to \(\boldsymbol{c}(\boldsymbol{r},\boldsymbol{r},\omega)\) when \(\rm{Im}[\varepsilon(\omega)]\neq0\).  This divergence stems from an assumption that the current flowing within the dielectric can have an arbitrarily rapid variation in space, as can be seen from the equal time correlation function~\cite{phi10},
\begin{equation}
	\langle0|\hat{\boldsymbol{j}}(\boldsymbol{r},t)\boldsymbol{\otimes}\hat{\boldsymbol{j}}(\boldsymbol{r}^{\prime},t)|0\rangle=\hbar\pi\boldsymbol{\mathds{1}}_{3}\delta^{(3)}(\boldsymbol{r}-\boldsymbol{r}^{\prime})\int_{0}^{\infty}d\omega\omega^{2}{\rm{Im}}[\varepsilon(\omega)], \label{current-corr}
\end{equation}
where the delta function on the right hand side of (\ref{current-corr}) shows that this current is made up from arbitrarily large component wavevectors.  The divergence would therefore be removed if we assumed that electromagnetic energy is always absorbed over some non-zero volume of space, replacing \(\boldsymbol{\mathds{1}}_{3}\delta^{(3)}(\boldsymbol{r}-\boldsymbol{r}^{\prime}){\rm{Im}}[\varepsilon(\omega)]\) in (\ref{current-corr}) with \(\rm{Im}[\boldsymbol{\varepsilon}(\boldsymbol{r},\boldsymbol{r}^{\prime},\omega)]\).  Once this description is adopted then it is possible to obtain finite results for \(\boldsymbol{c}(\boldsymbol{r},\boldsymbol{r},\omega)\).
\par
The link between such divergences encountered in macroscopic QED and the failure to account for spatial dispersion has been previously alluded to (e.g.~\cite{nar10,ruk61}), but to the authors' knowledge there has been no attempt to incorporate spatial dispersion into macroscopic QED from the outset to eliminate these divergences.  Perhaps part of the reason for this is that one requires a knowledge of the entire wavevector dependence of the permittivity of material samples, and such a characterisation is currently unavailable.  Both the theoretical and experimental understanding of spatial dispersion are quite underdeveloped compared to other aspects of macroscopic electromagnetism, and there is much scope in future years for important developments in this topic.
\par
The structure of this paper is as follows: the standard description of spatial dispersion in a homogeneous medium is outlined in section~\ref{sec:spatial}. In section~\ref{sec:act} we give a Lagrangian whose field equations are the macroscopic Maxwell equations in a homogeneous medium with nonlocal response. This allows the classical theory to be canonically quantized (section~\ref{sec:quantization}) and the Hamiltonian is diagonalized in section~\ref{sec:diag}. We apply the resulting quantum description to thermal and zero-point fields in a homogeneous medium and show that the resulting electromagnetic field intensities are no longer infinite when we assume a particular model of spatial dispersion (section~\ref{sec:thermal}). In section~\ref{sec:Casimir} we describe a range of basic questions concerning the Casimir effect (zero-point and thermal electromagnetic fields) that cannot be tackled, even approximately, if spatial dispersion is not included.

\section{Spatial dispersion in macroscopic electromagnetism \label{sec:spatial}}
\par
In this section we briefly recall the form of the classical macroscopic Maxwell equations when spatial dispersion is included.  A detailed account can be found in~\cite{LLcm,ruk61,mel06}.  The electromagnetic response of all real materials is nonlocal in space as well as in time, and the $\bi{D}$ and $\bi{H}$ fields must take the form
\begin{eqnarray}
\bi{D}(\bi{r},t)=\varepsilon_0\int_{-\infty}^t\rmd t' \int\rmd^3\bi{r'} \, \bvarepsilon(\bi{r},\bi{r}',t-t')\bdot\bi{E}(\bi{r}',t'),  \label{Dgen} \\
\bi{B}(\bi{r},t)=\mu_0\int_{-\infty}^t\rmd t' \int\rmd^3\bi{r'} \, \bmu(\bi{r},\bi{r}',t-t')\bdot\bi{H}(\bi{r}',t'),   \label{Bgen}
\end{eqnarray}
where $\boldsymbol{\varepsilon}(\bi{r},\bi{r}',t-t')$ is the relative permittivity and $\boldsymbol{\mu}(\bi{r},\bi{r}',t-t')$ is the relative permeability.  If the system as a whole exhibits time-reversal symmetry (i.e the medium is neither in motion, nor is there an external magnetic field), then the dielectric functions obey the symmetry relation~\cite{LLcm}
\begin{equation}   
\fl
\varepsilon_{ij}(\bi{r},\bi{r}',t-t')=\varepsilon_{ji}(\bi{r}',\bi{r},t-t'),  \qquad \mu_{ij}(\bi{r},\bi{r}',t-t')=\mu_{ji}(\bi{r}',\bi{r},t-t').   \label{epsym}
\end{equation}
\par
Taking the simplest case first, we specialize to a homogeneous medium, and assume that the dielectric functions in  (\ref{Dgen}) and (\ref{Bgen}) depend only on the difference between the coordinates, $\bi{r}-\bi{r}'$. In frequency and wavevector space the dielectric functions then depend on a single wavevector, \(\boldsymbol{k}\) and are complex functions obeying~\cite{LLcm,ruk61}
\begin{equation}  \label{epcep}
\varepsilon_{ij}^*(\bi{k},\omega)=\varepsilon_{ij}(-\bi{k},-\omega),  \qquad \mu_{ij}^*(\bi{k},\omega)=\mu_{ij}(-\bi{k},-\omega).
\end{equation}
In terms of the wavevector, the symmetry relation (\ref{epsym}) can also be written as
\begin{equation}
	\varepsilon_{ij}(\bi{k},\omega)=\varepsilon_{ji}(-\bi{k},\omega),  \qquad \mu_{ij}(\bi{k},\omega)=\mu_{ji}(-\bi{k},\omega).   \label{epsymk}
\end{equation}
\par
	To further simplify the situation, we assume an isotopic dielectric with a centre of symmetry, where the permittivity takes the form~\cite{ruk61,LLcm}
\begin{equation}  \label{permhom}
\bvarepsilon(\bi{k},\omega)=\varepsilon_\perp(k,\omega)\left(\mathds{1}-\frac{\bi{k}\otimes\bi{k}}{k^2}\right)+\varepsilon_\parallel(k,\omega)\frac{\bi{k}\otimes\bi{k}}{k^2},
\end{equation}
where $\varepsilon_\perp(k,\omega)$ and $\varepsilon_\parallel(k,\omega)$ are the respectively referred to as the transverse and longitudinal permittivities, which depend only on the magnitude of the wavevector, \(k=|\boldsymbol{k}|\).  Note that the Kramers-Kronig relations still hold for $\varepsilon_\perp(k,\omega)$ and $\varepsilon_\parallel(k,\omega)$ at each value of $k$, as a consequence of the delayed time response in (\ref{Dgen}).
\par
	The electromagnetic response captured by the tensor permittivity (\ref{permhom}) can also be implemented by an $(\omega,k)$-dependent scalar permittivity and permeability~\cite{ruk61}. This is seen by confirming that the macroscopic Maxwell equations with permittivity (\ref{permhom}) ($\mu=1$) are identical to those with scalar $\varepsilon(k,\omega)$ and  $\mu(k,\omega)$ given by~\cite{ruk61}
\begin{equation}   \label{epmuequiv}
\varepsilon(k,\omega)=\varepsilon_\parallel(k,\omega), \qquad 1-\mu^{-1}(k,\omega)=\frac{\omega^2}{c^2k^2}\left[\varepsilon_\perp(k,\omega)-\varepsilon_\parallel(k,\omega)\right].
\end{equation}
The electromagnetic energy dissipated by the medium naturally splits into absorption of transverse fields and longitudinal fields~\cite{ruk61}; this means that the transverse and longitudinal permittivities independently contribute to the absorption properties of the medium. It then follows that in a dissipative medium we must have~\cite{ruk61}
\begin{equation}   \label{discon}
\mathrm{Im}\varepsilon_\perp(k,\omega)>0, \qquad \mathrm{Im}\varepsilon_\parallel(k,\omega)>0.
\end{equation}
Interestingly, the same dissipative medium, when described by $\varepsilon(k,\omega)$ and  $\mu(k,\omega)$ in (\ref{epmuequiv}), does not in general have $\mathrm{Im}\mu(k,\omega)>0$~\cite{ruk61}. This is clear from imposing (\ref{discon}) on (\ref{epmuequiv}), which gives $\mathrm{Im}\varepsilon(k,\omega)>0$ but does not constrain the sign of $\mathrm{Im}\mu(k,\omega)$. The imaginary part of $\mu(k,\omega)$ can be positive or negative in a dissipative medium because $\varepsilon(k,\omega)$ and  $\mu(k,\omega)$ do not govern the absorption of linearly independent components of the electromagnetic fields, in contrast to $\varepsilon_\perp(k,\omega)$ and $\varepsilon_\parallel(k,\omega)$.
\par
In what follows we present the canonical quantization of a homogeneous, isotropic medium described by the permittivity (\ref{permhom}). This is the simplest case, and the theory can be readily generalized to more complicated nonlocal dielectric functions.

\section{Action and field equations for electromagnetism in a spatially dispersive, homogeneous and isotropic medium \label{sec:act}}  
\par
Following a similar procedure to~\cite{phi10,hut92}, the quantum theory of electromagnetism in a spatially dispersive medium is constructed through initially finding an action, \(S\), whose equations of motion are the macroscopic Maxwell equations with the constitutive relations (\ref{Dgen}) and (\ref{Bgen}).
\par
The required expression for \(S\) is a simple generalization of the action for the macroscopic Maxwell equations in the absence of spatial dispersion~\cite{phi10}:  \(S\) remains a functional of the electromagnetic potentials  $\phi(\bi{r},t)$ and $\bi{A}(\bi{r},t)$, and a continuum of reservoir fields $\bi{X}_\omega(\bi{r},t)$.  The reservoir fields---although unobservable---are a necessary part of the description, for they allow the absorption of electromagnetic radiation to be described within a closed system.  Nevertheless, the reservoir ought not to be confused with the real microscopic degrees of freedom of the medium, and we emphasize that the classical theory developed in this section will contain no more information than we can glean from the macroscopic Maxwell equations.  
\par
To incorporate spatial dispersion into the picture, we use a \emph{nonlocal} coupling between the electromagnetic field and the reservoir fields, finding the required action to be of the form 
\begin{equation} \label{S}
S[\phi,\bi{A},\bi{X}_\omega]=S_{\mathrm{em}}[\phi,\bi{A}]+S_ \mathrm {X}[\bi{X}_\omega]+S_{\mathrm{int}}[\phi,\bi{A},\bi{X}_\omega],
\end{equation}
where $S_{\mathrm{em}}$ is the free electromagnetic action
\begin{equation} \label{Sem}
S_{\mathrm{em}}[\phi,\bi{A}]=\frac{\varepsilon_{0}}{2}\int\rmd^4 x\left(\bi{E}^{2}-c^{2}\bi{B}^{2}\right),
\end{equation}
where the fields are related to the potentials by \(\bi{E}=-\bnabla\phi-\partial_t\bi{A}\) and \(\bi{B}=\bnabla\boldsymbol{\times}\bi{A}\).  The quantity $S_ \mathrm{X}$ is the action for the free reservoir oscillators,
\begin{equation}
S_ \mathrm{X}[\bi{X}_\omega]=\frac{1}{2}\int\rmd^4 x\int_0^\infty\rmd\omega\left[(\partial_t\bi{X}_\omega)^{2}-\omega^2\bi{X}_\omega^{2}\right], 
\end{equation}
and $S_{\mathrm{int}}$ is the interaction term, coupling the electromagnetic fields to the reservoir,
\begin{eqnarray}
\fl
S_{\mathrm{int}}[\phi,\bi{A},\bi{X}_\omega]=\int_{-\infty}^\infty\rmd t \int\rmd^3 \bi{r} \int\rmd^3 \bi{r'} \int_0^\infty\rmd\omega\, \bi{X}_\omega(\bi{r},t)\bdot\bi{F}(\bi{r}-\bi{r'},\omega)\bdot\bi{E}(\bi{r'},t).  \label{Sint}
\end{eqnarray}
The interaction (\ref{Sint}) features a spatially nonlocal coupling bi-tensor $\bi{F}(\bi{r}-\bi{r'},\omega)$ that depends on two scalar functions $\alpha_1(|\bi{r}-\bi{r'}|,\omega)$ and $\alpha_2(|\bi{r}-\bi{r'}|,\omega)$,
\begin{equation}
	\bi{F}(\bi{r}-\bi{r'},\omega)=\alpha_1(|\bi{r}-\bi{r'}|,\omega)\mathds{1} +\bnabla\alpha_2(|\bi{r}-\bi{r'}|,\omega)\otimes\stackrel{\leftarrow}{\bnabla'}.\label{F} 
\end{equation}
The scalar functions appearing in (\ref{F}) are defined in $\bi{k}$-space in terms of the imaginary parts of the transverse and longitudinal permittivities of an arbitrary homogeneous medium,
\begin{eqnarray}
\alpha_1(k,\omega)=\left[\frac{2\varepsilon_0}{\pi}\omega\,\mathrm{Im}\varepsilon_\perp(k,\omega)\right]^{1/2},  \label{al1}    \\[3pt]
 \alpha_2(k,\omega)=\frac{1}{k^2}\sqrt{\frac{2\varepsilon_0\omega}{\pi}}\left[\sqrt{\mathrm{Im}\varepsilon_\parallel(k,\omega)}-\sqrt{\mathrm{Im}\varepsilon_\perp(k,\omega)}\right].   \label{al2} 
\end{eqnarray}
The real parts of the permittivities appearing in (\ref{al1})--(\ref{al2}) are determined by the imaginary parts through the Kramers-Kronig relation~\cite{LLcm,jac}
\begin{equation}  \label{KK}
	\mathrm{Re}\,\varepsilon_{\{\perp,\parallel\}}(k,\omega')-1=\frac{2}{\pi}\mathrm{P}\int_0^\infty\rmd\omega\frac{\omega\,\mathrm{Im}\varepsilon_{\{\perp,\parallel\}}(k,\omega)}{\omega^2-\omega'^2},
\end{equation}
Therefore the appearance of only the imaginary parts of the dielectric function within (\ref{S}) should not be surprising, for we do not have the freedom to choose the real and imaginary parts of the susceptibilities independently.  As the permittivities depend on the magnitude of the distance (or wavevector), it is evident from (\ref{F}) that the nonlocal coupling tensor, \(\boldsymbol{F}\) possesses the symmetry, \(F_{ij}(\bi{r}-\bi{r'},\omega)=F_{ji}(\bi{r'}-\bi{r},\omega)\).  In the limiting case of a homogeneous medium lacking spatial dispersion, $\varepsilon_\perp(k,\omega)=\varepsilon_\parallel(k,\omega)=\varepsilon(\omega)$, the action (\ref{S})--(\ref{al2}) reduces to that in~\cite{phi10}, for the particular case where \(\mu=1\).
\par
We now show that the classical equations of motion derived from (\ref{S}) are the macroscopic Maxwell equations in a spatially dispersive medium.  Variation of the action with respect to the potentials $\phi$ and $\bi{A}$ leads to,
\begin{eqnarray}
\varepsilon_0\bnabla\bdot\bi{E}+\int_0^\infty\rmd\omega\int\rmd^3 \bi{r'}\,\bnabla\bdot\bi{F}(\bi{r}-\bi{r'},\omega)\bdot\bi{X}_\omega(\bi{r'},t)=0,   \label{Eeq} \\[3pt]
-\frac{1}{\mu_0}\bnabla\boldsymbol{\times}\bi{B}+\varepsilon_0\partial_t\bi{E}+\int_0^\infty\rmd\omega\int\rmd^3 \bi{r'}\,\bi{F}(\bi{r}-\bi{r'},\omega)\bdot\partial_t\bi{X}_\omega(\bi{r'},t)=0,  \label{Beq}
\end{eqnarray}
while variation with respect to $\bi{X}_\omega$ gives, 
\begin{equation}
-\partial_t^2\bi{X}_\omega-\omega^2\bi{X}_\omega+\int\rmd^3 \bi{r'}\,\bi{F}(\bi{r}-\bi{r'},\omega)\bdot\bi{E}(\bi{r'},t)=0.  \label{Xeq} 
\end{equation}
The equation of motion (\ref{Xeq}) is that of a field of driven harmonic oscillators, which has the general solution,
\begin{eqnarray}  
\bi{X}_\omega(\bi{k},\omega')=\frac{\bi{F}(\bi{k},\omega)\bdot\bi{E}(\bi{k},\omega')}{\omega^2-(\omega'+\rmi 0^+)^2}+\delta(\omega-\omega')\bi{Z}_\omega(\bi{k})+\delta(\omega+\omega')\bi{Z}^*_\omega(\bi{k})\nonumber\\
\label{Xeqsol}
\end{eqnarray}
The infinitesimal positive number $0^+$ in (\ref{Xeqsol}) amounts to a retarded boundary condition on the motion of the oscillator (i.e. the oscillator responds to the behavior of the electric field in the past), while $\bi{Z}_\omega(\bi{k})$ is an arbitrary function that, in the classical case, can be chosen to specify the configuration of the reservoir at some initial time, \(t=t_{0}\).
\par
The equations satisfied by the electromagnetic fields (\ref{Eeq}--\ref{Beq}) can be put in the desired form through applying the result (\ref{Xeqsol}). This is most easily carried out in the Fourier domain.  The Fourier-transformed versions of (\ref{Eeq}) and (\ref{Beq}) both contain the quantity $\int_0^\infty\rmd\omega\bi{F}(\bi{k},\omega)\bdot\bi{X}_\omega(\bi{k},\omega')$, which is evaluated using (\ref{Xeqsol}) as follows.  Firstly, it is clear from (\ref{F})--(\ref{al2}) that the square of the non--local coupling tensor is proportional to the dissipation in the medium,
\begin{equation}
\bi{F}(\bi{k},\omega)\bdot\bi{F}(\bi{k},\omega)=\frac{2\varepsilon_0\omega}{\pi}\mathrm{Im}\bvarepsilon(\bi{k},\omega),   \label{FF}
\end{equation}
where $\bvarepsilon(\bi{k},\omega)$ is the (tensor) permittivity (\ref{permhom}).  Rewriting the first term in (\ref{Xeqsol}) using the identity
\begin{equation}
\frac{1}{\omega^2-(\omega'+\rmi 0^+)^2}=\mathrm{P}\left(\frac{1}{\omega^2-\omega'^2}\right)+\frac{\rmi\pi}{2\omega}[\delta(\omega-\omega')-\delta(\omega+\omega')], \label{Ps}
\end{equation}
and employing (\ref{FF}) and the Kramers-Kronig relation (\ref{KK}), we find the required expression,
\begin{eqnarray}
\fl
\int_0^\infty\rmd\omega\bi{F}(\bi{k},\omega)\bdot\bi{X}_\omega(\bi{k},\omega')=\varepsilon_0\left[\bvarepsilon(\bi{k},\omega')-\mathds{1}\right]\bdot\bi{E}(\bi{k},\omega')+\bi{F}(\bi{k},\omega')\bdot\bi{Z}_{\omega'}(\bi{k}).  \label{FX}
\end{eqnarray}
Substituting this result (\ref{FX}) in the Fourier-transformed versions of (\ref{Eeq}) and (\ref{Beq}), we obtain the following form for the electromagnetic field equations
\begin{equation}
\rmi \bi{k}\bdot\bi{D}(\bi{k},\omega)=\sigma(\bi{k},\omega), \qquad \rmi \bi{k}\boldsymbol{\times}\bi{H}(\bi{k},\omega)=-\rmi\omega\bi{D}(\bi{k},\omega)+\bi{j}(\bi{k},\omega),  \label{macmax} 
\end{equation}
where the \(\boldsymbol{D}\) and \(\boldsymbol{H}\) fields are given by,
\begin{equation}
\bi{D}(\bi{k},\omega)=\varepsilon_0\bvarepsilon(\bi{k},\omega)\bdot\bi{E}(\bi{k},\omega),  \qquad \bi{H}(\bi{k},\omega)=\mu_0^{-1}\bi{B}(\bi{k},\omega),   \label{DHdef}
\end{equation}
and the charge and current densities are related to the initial configuration of the reservoir,
\begin{equation}
\sigma(\bi{k},\omega)=-\rmi \bi{k}\bdot\bi{F}(\bi{k},\omega)\bdot\bi{Z}_{\omega}(\bi{k}),  \qquad \bi{j}(\bi{k},\omega)=-\rmi\omega \bi{F}(\bi{k},\omega)\bdot\bi{Z}_{\omega}(\bi{k}).  \label{sigmaj}
\end{equation}
As anticipated, equations (\ref{macmax}) and (\ref{DHdef}) are the macroscopic Maxwell equations in a homogeneous, spatially dispersive medium with permittivity (\ref{permhom}) (the other two Maxwell equations are identities in terms of the electromagnetic potentials). It is evident from (\ref{sigmaj}) that the charge conservation law 
\begin{equation}  \label{conservation}
\partial_t\sigma(\bi{r},t)+ \bnabla\bdot \bi{j}(\bi{r},t)=0
\end{equation}
is identically true.
\par
Having shown that the action (\ref{S}) reproduces the classical theory of electromagnetism in a spatially dispersive medium, we are now in a position to develop the quantum theory.  Before doing so we give the expression for the electromagnetic Green function in the medium, which will be seen to determine the properties of the field operators in the quantum theory (see section~\ref{sec:diag}).  The electric field within the medium satisfies an inhomogeneous wave equation, which can be derived through substituting $\bi{B}(\bi{k},\omega)=\boldsymbol{k}\boldsymbol{\times}\bi{E}(\bi{k},\omega)/\omega$, into the second of (\ref{macmax}),
\begin{eqnarray}
\fl\left[\left( \frac{\omega^2}{c^2}\varepsilon_\perp(k,\omega)-k^2\right)\left(\mathds{1}-\frac{\bi{k}\otimes\bi{k}}{k^2}\right)+ \frac{\omega^2}{c^2k^2}\varepsilon_\parallel(k,\omega)\bi{k}\otimes\bi{k}              \right]\bdot\bi{E}(\bi{k},\omega)=-\rmi\mu_0\omega\bi{j}(\bi{k},\omega).\nonumber\\\label{E-field}
\end{eqnarray}
This has the solution
\begin{equation}   \label{Eksol}
\bi{E}(\bi{k},\omega)=\rmi\mu_0\omega\bi{G}(\bi{k},\omega)\bdot\bi{j}(\bi{k},\omega),
\end{equation}
where $\bi{G}(\bi{k},\omega)$ is the inverse of the matrix on the left hand side of (\ref{E-field}), known as the Green bi-tensor:
\begin{equation}
\fl
\bi{G}(\bi{k},\omega)=-\left[ \frac{\omega^2}{c^2}\varepsilon_\perp(k,\omega)-k^2\right]^{-1}\left(\mathds{1}-\frac{\bi{k}\otimes\bi{k}}{k^2}\right)- \frac{c^2}{\omega^2k^2\varepsilon_\parallel(k,\omega)}\bi{k}\otimes\bi{k}.   \label{Green}
\end{equation}
For reference, we note that the Green bi-tensor (\ref{Green}) satisfies
\begin{equation}   \label{Grel}
\frac{\omega^2}{c^2}\bi{G}^*(\bi{k},\omega) \bdot   \mathrm{Im}{\bvarepsilon}(\bi{k},\omega) \bdot \bi{G}(\bi{k},\omega)= \mathrm{Im}\bi{G}(\bi{k},\omega) .
\end{equation}

\section{Quantization\label{sec:quantization}}
\par
The quantization of the field theory given in (\ref{S})--(\ref{al2}) does not fundamentally differ from~\cite{phi10}.  The distinction between this theory and~\cite{phi10} is that here we have no coupling between the reservoir and the magnetic field, and the coupling of the reservoir to the electric field is spatially nonlocal.  The procedure is nevertheless the same: impose the canonical commutation relations between the fields and their associated canonical momenta, and construct the Hamiltonian operator from the action, (\ref{S}).
\par
The canonical momenta of the dynamical variables $\phi$, $\bi{A}$ and $\bi{X}_\omega$ are, respectively,
\begin{equation}
\fl
\mathit{\Pi}_\phi=0,  \quad     \bi{\Pi}_A=-\varepsilon_0\bi{E}-\int_0^\infty \rmd\omega\int\rmd^3 \bi{r'}\,\bi{F}(\bi{r}-\bi{r'},\omega)\bdot\bi{X}_\omega(\bi{r'},t),       \quad     \bi{\Pi}_{X_\omega}=\partial_t\bi{X}_\omega. \label{mom}
\end{equation}
Following the standard Coulomb-gauge formulation of quantum electrodynamics~\cite{wei}, we eliminate $\phi$ as a dynamical variable and impose the constraints
\begin{equation} \label{constraints}
\bnabla\bdot \bi{A}=0, \qquad \bnabla\bdot\bi{\Pi}_ A=0.
\end{equation}
With these constraints the equal--time commutation relation between $\boldsymbol{\hat{A}}$ and $\boldsymbol{\hat{\bi{\Pi}}}_{\!A}$ take the usual form~\cite{mel06,wei}
\begin{equation} \label{comem}
\left[ \boldsymbol{\hat{A}}(\bi{r},t),\boldsymbol{\hat{\bi{\Pi}}}_{\!A} (\bi{r'},t) \right]=\rmi\hbar \, \bdelta_{\mathrm{T}}(\mathbf{r}-\mathbf{r'}),
\end{equation}
where $\bdelta_{\mathrm{T}}(\bi{r}-\bi{r'})$ is the transverse delta function
\begin{equation}
\bdelta_{\mathrm{T}}(\bi{r}-\bi{r'})=\mathds{1}\delta(\bi{r}-\bi{r'})+\bnabla \otimes \bnabla \left(\frac{1}{4\pi|\bi{r}-\bi{r'}|}\right).
\end{equation} 
No such constraints are imposed on the reservoir, and we have the usual commutation relations
\begin{equation} \label{comX}
\left[\boldsymbol{\hat{X}}_{\omega}(\bi{r},t),\boldsymbol{\hat{\bi{\Pi}}}_{\!X_{\omega'}} (\bi{r'},t)\right]=\rmi\hbar\mathds{1}\delta(\omega-\omega')\delta(\bi{r}-\bi{r'}),
\end{equation}
The constraints (\ref{constraints}) give the following division of the electric-field operator into transverse and longitudinal parts, \(\boldsymbol{\hat{E}}=\boldsymbol{\hat{E}}_\mathrm{T}+\boldsymbol{\hat{E}}_\mathrm{L}\),
\begin{eqnarray}
\boldsymbol{\hat{E}}_\mathrm{T}&=-\partial_t\boldsymbol{\hat{A}}\nonumber\\
\boldsymbol{\hat{E}}_\mathrm{L}&=-\bnabla\hat{\phi}=-\frac{1}{\varepsilon_0}\left[\int_0^\infty \rmd\omega\int\rmd^3 \bi{r'}\,\bi{F}(\bi{r}-\bi{r'},\omega)\bdot\boldsymbol{\hat{X}}_\omega(\bi{r'},t)\right]_\mathrm{L}. \label{ETL}
\end{eqnarray}
\par
The construction of the Hamiltonian follows the steps explained in detail within~\cite{phi10} for a very similar action, with the result 
\begin{eqnarray}
\fl
\hat{H}=&\int\rmd^3\bi{r}\Bigg\{\frac{1}{2\varepsilon_0}\left[\boldsymbol{\hat{\bi{\Pi}}}_{\!A}+\int_0^\infty\! \rmd\omega\int\! \rmd^3 \bi{r'}\,\bi{F}(\bi{r}-\bi{r'},\omega)\bdot\boldsymbol{\hat{X}}_\omega(\bi{r'},t)\right]^{2}+\frac{1}{2\mu_0}(\bnabla\boldsymbol{\times}\boldsymbol{\hat{A}})^2 \nonumber \\[3pt]
\fl
 & +\frac{1}{2}\int_0^\infty\rmd\omega\left[\boldsymbol{\hat{\bi{\Pi}}}_{\!X_{\omega}}^2+\omega^2\boldsymbol{\hat{X}}_\omega^2\right]   
  \Bigg\}.  \label{H}
\end{eqnarray}
In the absence of any magnetic properties the Hamiltonian (\ref{H}) is of the same form as that in~\cite{phi10} with the replacement  \(\alpha(\boldsymbol{r},\omega)\hat{\boldsymbol{X}}_{\omega}(\boldsymbol{r})\to\int d\boldsymbol{r}^{\prime}\boldsymbol{F}(\boldsymbol{r}-\boldsymbol{r}^{\prime},\omega)\hat{\boldsymbol{X}}_{\omega}(\boldsymbol{r}^{\prime})\).  As one might expect, with the constraints (\ref{constraints}), the equations of motion for the operators are formally identical to the classical equations of motion (\ref{Eeq})--(\ref{Xeq}).  Due to the similarity of the form of the Hamiltonian, the demonstration of this is almost identical to that given in~\cite{phi10}.

\section{Diagonalization of the Hamiltonian} \label{sec:diag}
\par
The Hamiltonian (\ref{H}) is a quadratic combination of the field operators, and can be brought into the diagonal form,
\begin{equation} \label{Hdiag}
\hat{H}=\int\rmd^3 \bi{r}\int_0^\infty\rmd\omega\,\hbar\omega\boldsymbol{\hat{C}}^\dagger(\bi{r},\omega)\bdot\boldsymbol{\hat{C}}(\bi{r},\omega).
\end{equation}
This form of Hamiltonian amounts to the statement that the total energy of the system is equal to the integral over frequency of the number of quanta in the normal modes, weighted by the energy \(\hbar\omega\).  We may think of an excitation of each normal mode as something like a `polariton', which in this context would refer to a point--like excitation within the medium, and an associated electromagnetic field.  These point--like excitations may seem strange considering the discussion given in the introduction.  However the inclusion of spatial dispersion has the consequence that these excitations are now non--locally related to the current and charge density within the medium.  For instance, the state
 \(\hat{\boldsymbol{C}}^\dagger_{\omega}(\boldsymbol{r},t)|0\rangle\) now corresponds to an excitation of a current within the medium that is smeared out over a region of space centred at \(\boldsymbol{r}\). These `polaritons' are bosons, and the vector creation  $\boldsymbol{\hat{C}}^\dagger(\bi{r},\omega)$ and annihilation $\boldsymbol{\hat{C}}(\bi{r},\omega)$ operators therefore obey the commutation relations
\begin{equation} \label{CC}
\left[\boldsymbol{\hat{C}}(\bi{r},\omega),\boldsymbol{\hat{C}}^\dagger(\bi{r'},\omega')\right]=\mathds{1} \delta(\omega-\omega') \delta(\bi{r}-\bi{r'}).
\end{equation}
The time-dependence of these operators is given by
\begin{equation} \label{Ct}
\boldsymbol{\hat{C}}(\bi{r},t,\omega)=\boldsymbol{\hat{C}}(\bi{r},\omega)e^{-\rmi\omega t},
\end{equation}
which is consistent with (\ref{Hdiag})--(\ref{CC}).  The diagonalization procedure leading to (\ref{Hdiag}) is described in detail in~\cite{phi10} (more general versions of the procedure, required for microscopic models of dielectrics, are given in~\cite{hut92,sut04b}). Given the similarity to~\cite{phi10}, we confine ourselves to a sketch of the argument.
\par
The transformation of (\ref{H}) into (\ref{Hdiag}) is achieved through a canonical transformation of the fields, \(\hat{\boldsymbol{A}}\), \(\hat{\boldsymbol{\Pi}}_{\boldsymbol{A}}\), \(\hat{\boldsymbol{X}}_{\omega}\) \& \(\hat{\boldsymbol{\Pi}}_{\boldsymbol{X}_{\omega}}\).  In general this transformation can be written in the form,
\begin{eqnarray} 
\boldsymbol{\hat{A}}(\bi{r},t)&=\int\rmd^3 \bi{r'}\int_0^\infty\rmd\omega\, [\bi{f}_A(\bi{r}-\bi{r'},\omega)\bdot\boldsymbol{\hat{C}}(\bi{r'},t,\omega)+\mbox{h.c.}],   \label{AC} \\[3pt]
\boldsymbol{\hat{\bi{\Pi}}}_{\!A}(\bi{r},t)&=\int\rmd^3 \bi{r'}\int_0^\infty\rmd\omega\, [\bi{f}_{\mathit{\Pi}_A}(\bi{r}-\bi{r'},\omega)\bdot\boldsymbol{\hat{C}}(\bi{r'},t,\omega)+\mbox{h.c.}],  \label{PAC}   \\[3pt]
\boldsymbol{\hat{X}}_\omega(\bi{r},t)&=\int\rmd^3 \bi{r'}\int_0^\infty\rmd\omega^{\prime}\, [\bi{f}_{X}(\bi{r}-\bi{r'},\omega,\omega')\bdot\boldsymbol{\hat{C}}(\bi{r'},t,\omega')+\mbox{h.c.}],   \label{XC} \\[3pt]
\boldsymbol{\hat{\bi{\Pi}}}_{\!X_{\omega}}(\bi{r},t)&=\int\rmd^3 \bi{r'}\int_0^\infty\rmd\omega^{\prime}\, [\bi{f}_{\mathit{\Pi}_X}(\bi{r}-\bi{r'},\omega,\omega')\bdot\boldsymbol{\hat{C}}(\bi{r'},t,\omega')+\mbox{h.c.}],  \label{PXC}  
\end{eqnarray}
where the $\boldsymbol{f}_{\{A,\Pi_{A},X,\Pi_{X}\}}$ are c-number bi-tensors. It is also useful to define the bi-tensor $\bi{f}_E$ that gives the expansion of the electric-field operator, \(\hat{\boldsymbol{E}}\) in terms of the $\boldsymbol{\hat{C}}^\dagger(\bi{r},\omega)$  and $\boldsymbol{\hat{C}}(\bi{r},\omega)$ operators.  Inserting (\ref{PAC}) and (\ref{XC}) into our earlier expression for the electric field operator (\ref{mom}), we find this to be
\begin{equation}  \label{fE}
\fl
\bi{f}_E(\bi{r}-\bi{r'},\omega)=-\frac{1}{\varepsilon_0}\bi{f}_{\mathit{\Pi}_A}(\bi{r}-\bi{r'},\omega)-\frac{1}{\varepsilon_0}\int_0^\infty \! \rmd\omega'  \! \! \int  \! \rmd^3 \bi{r''}\,\bi{F}(\bi{r}-\bi{r''},\omega')\bdot\bi{f}_{X}(\bi{r''}-\bi{r'},\omega',\omega).
\end{equation}
The commutation relations satisfied by the various operators imply that the inverse of the transformation (\ref{AC})--(\ref{PXC}) is given by
\begin{eqnarray}
\fl
\boldsymbol{\hat{C}}(\bi{r},t,\omega)=-\frac{\rmi}{\hbar}\int\rmd^3 \bi{r'}\left\{\boldsymbol{\hat{A}}(\bi{r'},t)\bdot\bi{f}^*_{\mathit{\Pi}_A}(\bi{r'}-\bi{r},\omega)-\boldsymbol{\hat{\bi{\Pi}}}_{\!A}(\bi{r'},t)\bdot\bi{f}^*_{A}(\bi{r'}-\bi{r},\omega) \right. \nonumber  \\[3pt]
\fl
 \qquad\  \left. +\int_0^\infty \rmd\omega'\left[\boldsymbol{\hat{X}}_{\omega'}(\bi{r'},t)\bdot\bi{f}^*_{\mathit{\Pi}_X}(\bi{r'}-\bi{r},\omega',\omega)-\boldsymbol{\hat{\bi{\Pi}}}_{\!X_{\omega'}}(\bi{r'},t)\bdot\bi{f}^*_{X}(\bi{r'}-\bi{r},\omega',\omega)\right]  \right\}  . \label{Ccan}
 \end{eqnarray}
It must now be demonstrated that it is possible to find a transformation (\ref{AC})--(\ref{PXC}) that will turn (\ref{H}) into (\ref{Hdiag}), without disturbing the canonical commutation relations.  This transformation should also be such that (\ref{Ccan}) is consistent with (\ref{AC})--(\ref{PXC}).  We do this briefly in words---a more rigorous argument can be found in~\cite{phi10}.
\par
Given that the field operators (\ref{AC})--(\ref{PXC}) formally satisfy the classical equations of motion (\ref{Eeq})--(\ref{Xeq}), and the \(\hat{\boldsymbol{C}}_{\omega}\) have an assumed time dependence given by (\ref{Ct}), the transformation can only work if there is a direct relation between the solutions to the classical equations of motion in the frequency domain and the \(\boldsymbol{f}_{\{A,\Pi_{A},X,\Pi_{X}\}}\).  For example, the frequency-domain expression for the classical electric field is given by the Fourier transform of (\ref{Eksol}) with respect to \(\boldsymbol{k}\).  A comparison between the operator expression for \(\hat{\boldsymbol{E}}\) and its classical counterpart then shows that the \(\hat{\boldsymbol{C}}_{\omega}\) operator is analogous to the classical quantity \(\boldsymbol{Z}_{\omega}e^{-i\omega t}\) introduced in (\ref{Xeqsol}).  One can therefore replace \(\boldsymbol{Z}_{\omega}e^{-i\omega t}\) with \(\zeta(\omega)\hat{\boldsymbol{C}}_{\omega}\) in the solution to the classical problem and thereby obtain (\ref{AC})--(\ref{PXC}): having done this, the operators will satisfy the classical equations of motion, and this will be automatically consistent with (\ref{Hdiag}).  The only quantity left to determine is \(\zeta(\omega)\), which is a function of frequency chosen so that the commutation relations (\ref{comem}), (\ref{comX}), and (\ref{CC}) are all simultaneously satisfied.  The necessary analysis can be found in~\cite{sch08}, where it is determined that \(\zeta(\omega)=2\pi\sqrt{\hbar/2\omega}\).  The quantity \(\boldsymbol{f}_{E}\) is then equal to
\begin{equation}  \label{fEsol}
	\bi{f}_E(\bi{k},\omega)=\mu_0\omega\sqrt{\frac{\hbar\omega}{2}}\,\bi{G}(\bi{k},\omega)\bdot\bi{F}(\bi{k},\omega).
\end{equation}
Similarly the bi-tensorial coefficients associated with the electromagnetic operators are,
\begin{eqnarray}
\bi{f}_A(\bi{k},\omega)=-\frac{\rmi}{\omega}\left[\bi{f}_E(\bi{k},\omega)\right]_\mathrm{T},   \label{fAsol}  \\
\bi{f}_{\mathit{\Pi}_A}(\bi{k},\omega)= -\varepsilon_0\boldsymbol{\varepsilon}(\bi{k},\omega)\bdot\bi{f}_E(\bi{k},\omega)-\sqrt{\frac{\hbar}{2\omega}}\,\bi{F}(\bi{k},\omega)  \label{fPAsol}
\end{eqnarray}
and those of the reservoir are,
\begin{eqnarray}
\bi{f}_{X}(\bi{k},\omega',\omega)= \frac{\rmi}{\omega}\bi{f}_{\mathit{\Pi}_X}(\bi{k},\omega',\omega),  \label{fXsol}  \\
\bi{f}_{\mathit{\Pi}_X}(\bi{k},\omega',\omega)=   -\frac{\rmi\omega}{2\omega'}\left[ \frac{1}{\omega'-\omega-\rmi 0^+}  + \frac{1}{\omega'+\omega}  \right]\bi{F}(\bi{k},\omega')\bdot\bi{f}_E(\bi{k},\omega)   \nonumber   \\
\qquad\qquad\qquad\ \   -\rmi\sqrt{\frac{\hbar\omega}{2}}\,\mathds{1}\delta(\omega-\omega').  \label{fPXsol}
\end{eqnarray}
The external charge and current densities (\ref{sigmaj}) appearing in the classical problem are now operators,
\begin{eqnarray}  \label{jop}
\hat{\sigma}(\bi{k},\omega)=-2\pi\rmi\left[\frac{\hbar\varepsilon_0}{\pi}\,\mathrm{Im}\varepsilon_\parallel(k,\omega)\right]^{1/2}\bi{k}\bdot\boldsymbol{\hat{C}}(\bi{k},\omega)\\
\boldsymbol{\hat{\jmath}}(\bi{k},\omega)=-2\pi\rmi\sqrt{\frac{\hbar\omega}{2}}\,\bi{F}(\bi{k},\omega) \bdot\boldsymbol{\hat{C}}(\bi{k},\omega)
\end{eqnarray}
From (\ref{CC}), the $\bi{k}$-space creation and annihilation operators have the commutation relations
\begin{equation} \label{CCk}
\left[\boldsymbol{\hat{C}}(\bi{k},\omega),\boldsymbol{\hat{C}}^\dagger(\bi{k'},\omega')\right]=(2\pi)^3\mathds{1} \delta(\omega-\omega') \delta(\bi{k}-\bi{k'}), 
\end{equation}
so that the charge and current operators (\ref{jop}) satisfy,
\begin{eqnarray} 
\left[\hat{\sigma}(\bi{k},\omega),\hat{\sigma}^\dagger(\bi{k'},\omega')\right]&=(2\pi)^5\frac{\hbar k^{2}\varepsilon_0}{\pi}\mathrm{Im}\varepsilon_{\parallel}(k,\omega)\delta(\omega-\omega') \delta(\bi{k}-\bi{k'})\label{sscom1}\\
\left[\boldsymbol{\hat{\jmath}}(\bi{k},\omega),\boldsymbol{\hat{\jmath}}^\dagger(\bi{k'},\omega')\right]&=(2\pi)^5\frac{\hbar\omega^2\varepsilon_0}{\pi}\mathrm{Im}\boldsymbol{\varepsilon}(\bi{k},\omega)\delta(\omega-\omega') \delta(\bi{k}-\bi{k'}),   \label{jjcom1}  
\end{eqnarray}
where we have applied (\ref{FF}).  It is interesting that, without any dependence of \(\varepsilon_{\parallel}\) on \(k\), (\ref{sscom1}) implies that the correlation between the charge density within the medium grows with \(k\).  When \(\rm{Im}[\boldsymbol{\varepsilon}]\) depends on \(\boldsymbol{k}\) such that it tends to zero as \(k\to\infty\), then the commutation relation (\ref{jjcom1}) will modify the correlation function for the current density from the delta function given in (\ref{current-corr}) to a correlation that is spread out over a finite region of space.  This is the anticipated effect of spatial dispersion discussed in the introduction.

\section{Thermal and zero-point field correlations in a medium with spatial dispersion\label{sec:thermal}}
\par
We are now in a position to return to the problem raised in the introduction.  It was shown that the arbitrarily localized current densities appearing in the local theory of macroscopic quantum electromagnetism give rise to divergent results.  Here we explicitly demonstrate that the introduction of spatial dispersion allows us to obtain finite results.  Our attention is confined to the equilibrium properties of the electromagnetic field within a homogeneous medium.
\par
In thermal equilibrium, the `polaritons' identified through the diagonalization of the Hamiltonian in section~\ref{sec:diag}, have an occupation number given by the Planck distribution~\cite{phi11}
\begin{equation}  
\left\langle\boldsymbol{\hat{C}}^\dagger(\bi{r},\omega)\otimes\boldsymbol{\hat{C}}(\bi{r'},\omega')\right\rangle=\mathcal{N}(\omega)\mathds{1}\delta(\omega-\omega') \delta(\bi{r}-\bi{r'}),  \label{CdC}
\end{equation}
where
\begin{equation}   
   \mathcal{N}(\omega):=\frac{1}{e^{\hbar\omega/k_{B}T}-1}.  \label{N}
\end{equation}
so that the current operator (\ref{jop}) has the thermal correlation
\begin{equation} 
	\fl
	\left\langle\boldsymbol{\hat{\jmath}}^\dagger(\bi{k},\omega)\otimes\boldsymbol{\hat{\jmath}}(\bi{k'},\omega')\right\rangle=(2\pi)^5\mathcal{N}(\omega)\frac{\hbar\omega^2\varepsilon_0}{\pi}\mathrm{Im}\boldsymbol{\varepsilon}(\bi{k},\omega)\delta(\omega-\omega') \delta(\bi{k}-\bi{k}),  \label{jdjk}
 \label{jjdk}   
\end{equation}
where we have again applied (\ref{FF}).  It is now straightforward to use (\ref{CdC}) and (\ref{fEsol}) to compute the equal-time field correlation functions in thermal equilibrium.  With use of the Green-function relation (\ref{Grel}) the results take the same form as in the local theory,
\begin{eqnarray}
\fl
\left\langle\boldsymbol{\hat{E}} (\bi{r}, t)\otimes\boldsymbol{\hat{E}}(\bi{r'}, t)\right \rangle&=\frac{\hbar\mu_0}{\pi}\int_0^\infty\rmd\omega\,\omega^2\coth\left(\frac{\hbar\omega}{2k_B T}\right)\mathrm{Im}\bi{G}(\bi{r}-\bi{r'},\omega)\label{EE}\\
\fl
\left\langle\boldsymbol{\hat{B}} (\bi{r}, t)\otimes\boldsymbol{\hat{B}}(\bi{r'}, t)\right \rangle&=\frac{\hbar\mu_0}{\pi}\int_0^\infty\rmd\omega\,\coth\left(\frac{\hbar\omega}{2k_B T}\right)\bnabla\boldsymbol{\times}\mathrm{Im}\bi{G}(\bi{r}-\bi{r'},\omega)\boldsymbol{\times}\stackrel{\leftarrow}{\bnabla'}\label{BB}
\end{eqnarray}
where the notation $\boldsymbol{\times}\stackrel{\leftarrow}{\bnabla'}$ denotes a curl with respect to the right-hand index, i.e\ $\bi{V}(\bi{r})\boldsymbol{\times}\stackrel{\leftarrow}{\bnabla'}= \bnabla \boldsymbol{\times}\bi{V}(\bi{r})$ for a vector field $\bi{V}(\bi{r})$.
\par
In this non--local theory we again consider the intensity of the electric field at a fixed frequency, \(\boldsymbol{c}(\boldsymbol{r},\boldsymbol{r},\omega)\), which at \(T>0\rm{K}\) is, from (\ref{EE})
\begin{equation}
	\boldsymbol{c}(\boldsymbol{r},\boldsymbol{r},\omega)=\frac{\hbar\mu_{0}}{\pi}\omega^{2}\coth\left(\frac{\hbar\omega}{2 k_{B}T}\right)\lim_{\boldsymbol{r}\to\boldsymbol{r}^{\prime}}\rm{Im}[\boldsymbol{G}(\boldsymbol{r}-\boldsymbol{r}^{\prime},\omega)] .            \label{intensity}
\end{equation}
For want of established expressions for the nonlocal permittivity of a real medium, we examine the behavior of (\ref{intensity}) in a medium described by the permittivity
\begin{eqnarray}
	\varepsilon_{\perp}(\omega)&=1-\frac{A}{\omega(\omega+i\gamma)},   \label{ept}\\
	\varepsilon_{\parallel}(k,\omega)&=1-\frac{A}{\omega(\omega+i\gamma)-\beta^{2}k^{2}},   \label{epl}
\end{eqnarray}
where \(A\) and \(\beta\) are arbitrary constants, and \(\gamma>0\) governs the absorption into the medium.  The nonlocal permittivity defined by (\ref{ept}) and (\ref{epl}) is of the same form as the hydrodynamic Drude model (see e.g.~\cite{raz11}), and can also be considered a limiting case of the Hopfield model~\cite{hop63}.  In general both \(\varepsilon_{\perp}\) and \(\varepsilon_{\parallel}\) would depend on \(k\), but in these models of homogeneous media the non--locality is confined to \(\varepsilon_{\parallel}\).  The imaginary part of the Green function in the limit as \(\boldsymbol{r}\to\boldsymbol{r}^{\prime}\) is related to its spatial Fourier transform by
\begin{equation}
	\lim_{\boldsymbol{r}\to\boldsymbol{r}^{\prime}}\rm{Im}[\boldsymbol{G}(\boldsymbol{r}-\boldsymbol{r}^{\prime},\omega)]=\int\frac{d^{3}\boldsymbol{k}}{(2\pi)^{3}}\rm{Im}[\boldsymbol{G}(\boldsymbol{k},\omega)] .  \label{ldos}
\end{equation}
Inserting (\ref{Green}) into (\ref{ldos}) then gives,
\begin{equation}
	\fl\lim_{\boldsymbol{r}\to\boldsymbol{r}^{\prime}}{\rm{Im}}[\boldsymbol{G}(\boldsymbol{r}-\boldsymbol{r}^{\prime},\omega)]=\int\frac{d^{3}\boldsymbol{k}}{(2\pi)^{3}}\Bigg\{\frac{\left(\mathds{1}-\frac{\boldsymbol{k}\boldsymbol{\otimes}\boldsymbol{k}}{k^{2}}\right)\frac{\omega^{2}}{c^{2}}{\rm Im}[\varepsilon_{\perp}(\omega)]}{\left|\frac{\omega^{2}}{c^{2}}\varepsilon_{\perp}(\omega)-k^{2}\right|^{2}}+\frac{\boldsymbol{k}\boldsymbol{\otimes}\boldsymbol{k} c^{2}{\rm Im}[\varepsilon_{\parallel}(k,\omega)]}{\omega^{2}k^{2}|\varepsilon_{\parallel}(k,\omega)|^{2}}\Bigg\}  .  \label{Gres1}
\end{equation}
In a local description, where \(\varepsilon_{\parallel}\) is only frequency dependent, the integral over the longitudinal part of the integrand in (\ref{Gres1}) diverges.  However,
\begin{equation}
	{\rm Im}[\varepsilon_{\parallel}(k,\omega)]=\frac{A\omega\gamma}{(\omega^{2}-\beta^{2}k^{2})^{2}+\omega^{2}\gamma^{2}}
\end{equation}
tends to zero as \(k^{-4}\), which is sufficient to make (\ref{Gres1}) converge.  We now explicitly evaluate (\ref{Gres1}), first performing the integral over the angle of the \(\boldsymbol{k}\) vector,
\begin{equation}
	\int_{0}^{2\pi}\frac{d\phi_{k}}{2\pi}\int_{0}^{\pi}\frac{\sin(\theta_{k})d\theta_{k}}{2\pi}\boldsymbol{k}\boldsymbol{\otimes}\boldsymbol{k}=\frac{k^{2}}{3\pi}\mathds{1} ,  \label{angint}
\end{equation}
where \(\boldsymbol{k}=k[\sin(\theta_{k})\cos(\phi_{k})\hat{\boldsymbol{x}}+\sin(\theta_{k})\sin(\phi_{k})\hat{\boldsymbol{y}}+\cos(\theta_{k})\hat{\boldsymbol{z}}]\).  
The integrals over the magnitude of the wavevector \(k\) may be performed through extending the range of integration from \([0,\infty)\) to \((-\infty,\infty)\), closing the contour of integration in either the upper or lower half \(k\)-plane, and applying the residue theorem.  This gives
\begin{eqnarray*}
	\int_{-\infty}^{\infty}\frac{k^{2}dk}{4\pi^{2}}\frac{1}{\left|k^{2}-\frac{\omega^{2}}{c^{2}}\varepsilon_{\perp}(\omega)\right|^{2}}&=\frac{1}{4\pi}\frac{{\rm Re}[\sqrt{\varepsilon_{\perp}(\omega)}]}{\frac{\omega}{c}{\rm Im}[\varepsilon_{\perp}(\omega)]}  ,  \\
	\int_{-\infty}^{\infty}\frac{k^{2}dk}{4\pi^{2}}\frac{{\rm Im}[\varepsilon_{\parallel}(k,\omega)]}{|\varepsilon_{\parallel}(k,\omega)|^{2}}&=\frac{A}{4\pi\beta^{3}}{\rm Re}[\sqrt{\omega^{2}-A+\rmi\omega\gamma}],
\end{eqnarray*}
which finally leads to the following expression for the intensity of the electric field at a fixed frequency:
\begin{equation}
	\fl
	\boldsymbol{c}(\boldsymbol{r},\boldsymbol{r},\omega)=\frac{\hbar}{\pi\varepsilon_{0}}\coth\left(\frac{\hbar\omega}{2 k_{B}T}\right)\mathds{1}\Bigg\{\frac{\omega^{3}}{6\pi c^{3}}{\rm Re}[\sqrt{\varepsilon_{\perp}(\omega)}]+\frac{A}{12\pi\beta^{3}}{\rm Re}[\sqrt{\omega^{2}-A+\rmi\omega\gamma}]\Bigg\}  .  \label{Gres2}
\end{equation}
In the limit \(T\to0\rm{K}\), the first bracketed term is proportional to the transverse spontaneous emission rate, already found in previous studies of quantum electromagnetism in absorbing media (e.g.~\cite{bar96}).  Meanwhile the second term---proportional to the longitudinal spontaneous emission rate---arises from the non--locality of the longitudinal permittivity, and diverges in the local limit \(\beta\to0\) as shown in (\ref{long-G}).  In the hydrodynamic Drude model, the polarizability of the medium is due to the motion of the conduction electrons, which are treated as a charged fluid.  The quantity \(\beta\) is related to a pressure term in the equations of motion for this fluid that serves to smooth out rapid variations in the density~\cite{raz11}.  Equation (\ref{Gres2}) shows that when the electromagnetic field is coupled to such a model medium and quantized, then this pressure term naturally regulates the divergence of the electric field intensity at a fixed frequency.  Furthermore, if we make the replacement, \(\coth(\hbar\omega/2k_{B}T)\to\coth(\hbar\omega/2k_{B}T)-1\) in (\ref{Gres2}), then a finite result may be obtained for the \emph{purely thermal} contribution to the \emph{total} (rather than single frequency) equilibrium field intensity in a homogeneous medium (\ref{EE}).  It would be interesting if processes such as spontaneous emission act as a probe of the non--local properties of a medium through the longitudinal emission rate.

\section{Spatial dispersion and the Casimir effect}  \label{sec:Casimir}
An extension of our results to inhomogeneous materials, including the case of piece-wise homogeneous configurations, is of course required to treat most problems of interest. In particular, extension of the thermal and zero-point results to materials with boundaries will give a full treatment of the Casimir effect that incorporates spatial dispersion. In this section we describe some fundamental problems in Casimir theory that can be tackled with a full account of spatial dispersion in real materials. Our purpose is to point out the deep significance of spatial dispersion for the Casimir effect, as this significance is not widely recognized in the literature. 

The Lifshitz formula~\cite{lif55} for the Casimir force between parallel half-spaces does not take account of spatial dispersion. The change in the force due to the nonlocal response of conduction electrons in a metal has been calculated in~\cite{esq04,esq06} and found to be small for current experimental regimes. Although the contribution of spatial dispersion is minor for Casimir forces between separate objects (except at very small separations), it is dominant for some components of the Casimir stress-energy tensor at the boundaries of those objects. In particular, if spatial dispersion is neglected, both the Casimir energy density and the lateral Casimir stress diverge at planar boundaries. This fact was noted in~\cite{sop02,bar05} for zero-point fields, but the divergence is present also for the purely thermal stress-energy even if zero-point fields are dropped. The divergent result for the thermal energy density at material boundaries, when only the frequency dependence of the permittivity is included, is well known in surface science (see the review~\cite{jou05}, and in particular equation (48) therein, which diverges at the surface). The behaviour of purely thermal radiation shows that the divergences in the Casimir stress-energy at material boundaries are not due to zero-point radiation specifically and therefore they are not related to the regularization of zero-point fields. The divergences are the result of integrating over evanescent fields on the boundary with arbitrarily large lateral wavevectors; the divergence occurs if the permittivity is taken to be independent of the wavevector, which is clearly an incorrect assumption when wavevectors up to infinite values are crucial in the calculation. A full treatment of the problem will require detailed knowledge of the wavevector dependence of the permittivities of real material samples. 

Divergences in the Casimir stress-energy tensor also occur for curved material boundaries when spatial dispersion is ignored. There is a distorting pressure on curved boundaries due to thermal and zero-point fields and a calculation of this pressure requires the correct (finite) component of the Casimir stress tensor perpendicular to the boundary. The perpendicular Casimir stress vanishes for flat boundaries but it diverges for curved boundaries if the permittivity is taken to depend only on frequency. This diverging distortion force, or self-force, is most familiar in the case of a dielectric ball and a spherical dielectric shell~\cite{mil80}. The result of Boyer~\cite{boy68} that the zero-point self-force on a perfectly conducting, infinitely thin spherical shell is directed outwards, has no direct physical relevance since such an object does not exist. Even in the case of Boyer's shell, however, the standard, experimentally supported, Lifshitz theory shows that the self-force is in fact inwardly directed and infinite, rather than outwardly directed and finite. All these self-force divergences are again present for purely thermal radiation so they are not zero-point phenomena per se. The divergences can again be traced to the contribution of fields with arbitrarily large wavevectors along the boundary, together with the assumption that the material properties do not change with wavevector. Using a cut-off at some large value of the lateral wavevector removes the divergence~\cite{can82} but would not be expected to give an accurate estimate of Casimir self-forces; instead the full details of the spatial dispersion of the object in question are needed.

Finally, the Casimir self-forces on inhomogeneous materials, where the permittivity changes continuously with position, also diverge when spatial dispersion is ignored~\cite{phi10a}. This fact shows that the divergences at sharp boundaries, mentioned above, are not removed if the boundary is smoothed so that the permittivity decreases continuously to one~\cite{phi10a}. In summary, a better understanding of the nonlocal response of real materials is required to address some basic problems in the theory of the Casimir effect, and these problems can also be posed for purely thermal radiation.

\section{Conclusions}
\par
We have shown that it is possible to formulate consistently a canonical theory of quantum electromagnetism in nonlocal dielectric media.  Although we specialised to the case of homogeneous and isotropic media, it is straightforward to extend (\ref{S}) to treat the general case.
\par
Our treatment of spatial dispersion was motivated by a basic demonstration that the local theory of macroscopic quantum electromagnetism can fail to yield finite predictions when applied within bulk media.  When a model for spatial dispersion is adopted such as the hydrodynamic Drude model, or the Hopfield model, then finite results for field intensities can be obtained (section~\ref{sec:thermal}).  In particular, it is interesting that in the local theory the purely thermal contribution to the mean field intensity is predicted to be infinite within a homogeneous medium, a result that becomes finite in our treatment.  Despite these encouraging results, further work is required to determine which model is appropriate for describing the nonlocal behaviour of a given material sample.
\par
We have pointed out that some well-known divergences that arise in Casimir theory are due to the neglect of nonlocal response (section~\ref{sec:Casimir}). A realistic treatment of non-locality in materials is thus essential for an accurate estimate of distortion-forces on dielectric bodies due to thermal and zero-point fields. 

\ack
We acknowledge financial support from EPSRC under Program Grant EP/I034548/1.


\end{document}